\documentclass[%
 reprint,
 floatfix,
 amsmath,amssymb,
 aps,
 pra,
]{revtex4-2}

\usepackage{graphicx}
\usepackage{dcolumn}
\usepackage{bm}
\usepackage{color}
\usepackage{silence}
\usepackage{physics}
\usepackage{comment}
\usepackage{ulem}
\usepackage{hyperref}
\begin{document}

\title{Purification of a monitored qubit: exact path‑integral solution}

\author{Matheus M. R. Poltronieri Martins $^1$ and Henrique Santos Lima$^1$}

\affiliation{ $^1$ Centro Brasileiro de Pesquisas Físicas, Rua Xavier Sigaud 150, Rio de Janeiro 22290-180, Brazil }

\date{\today}

\begin{abstract}
We investigate the purification dynamics of a single qubit under continuous-in-time monitoring. By employing a collisional model framework where the system interacts sequentially with ancillary qubits, we describe the conditioned evolution of the density matrix through a stochastic master equation. We show that for initial mixed states, the dynamics reduce to a multiplicative Langevin equation for a single scalar parameter representing the state's purity. This stochastic process is solved exactly using the Onsager--Machlup path-integral formalism, allowing us to derive the full probability distribution for the qubit's trajectories. Our analytical results reveal that purification is characterized by a dynamical crossover from a diffusion-dominated regime to a measurement-dominated regime, visible in the emergence of a bimodal state distribution. The analytical solutions are in strong agreement with numerical simulations, providing a robust theoretical benchmark for the study of information extraction in monitored quantum systems.
\end{abstract}

\maketitle

\section*{Introduction}

Quantum dynamics is shaped by the competition between unitary evolution and measurement. While unitary dynamics generate entanglement and scramble local information, measurements act to suppress coherence and project the system onto definite states. In hybrid quantum circuits, this competition can drive measurement-induced phase transitions (MIPT), separating volume-law entangled phases from area-law phases stabilized by sufficiently frequent monitoring \cite{LiChenFischer2018,Nahum,LiChenFisher2019_Hybrid_QC,gullans2020dynamical,Vasseur}. Related noisy monitored multi-qubit dynamics were analyzed by Schomerus, who emphasized how measurement noise modifies ergodic quantum evolution \cite{Schomerus2022}.

In ergodic many-body systems, the eigenstate thermalization hypothesis explains how isolated systems act as their own thermal baths, making local information inaccessible at late times \cite{JMdeutsch1991,Srednicki1994,rigol2008thermalization}. By contrast, many-body localized systems can retain memory of their initial conditions over long times \cite{Nandkishore_2015}. These transitions and their associated scaling laws have been extensively studied through percolation mappings, tensor-network descriptions, and conformal-field-theory structures \cite{grimmett1999percolation,VasseurPotterYou_tensornetwork_2019,Potter_2022}.

Purification dynamics provides a fundamental way to quantify the effect of monitoring in quantum systems. In many-body settings, the time scale required for a system to reach a pure state depends on the interplay between intrinsic scrambling and the measurement rate \cite{JacobsSteck,wiseman2010quantum,gullans2020dynamical}. In particular, Jacobs showed that feedback can accelerate qubit projection, highlighting purification as a controllable dynamical resource \cite{Jacobs2003}. Strong monitoring rapidly projects the state, leading to fast purification, while weak monitoring allows the dynamics to retain a mixed character for longer periods. Understanding the statistical properties of these purification trajectories is crucial for characterizing the steady states of monitored quantum matter.

In the present work, we shift our focus from the complexities of the many-body problem to the fundamental case of a single qubit under continuous monitoring. Unlike the discrete, instantaneous jumps associated with projective measurements, continuous monitoring represents a gradual state update driven by a steady stream of weak interactions \cite{Korotkov2001,Brun_2002,Wiseman_1996,Ciccarello_2022}. Korotkov provided an early continuous-measurement description of selective qubit evolution \cite{Korotkov2001}. We restrict our study to initial mixed states, for which the dynamics reduce to the evolution of diagonal density-matrix elements in the measurement basis.

The goal of this article is to derive a closed analytical solution for the purification of a continuously monitored qubit. We map the stochastic master equation to a Langevin equation with multiplicative noise and analyze the resulting dynamics using an Onsager--Machlup path-integral formulation \cite{Onsager53,Onsager53_2,Cugliandolo_2017_rules_of_calculus,review_Path_stochastic}. This approach allows us to obtain the full probability distributions for the state parameter and the purity. We validate our analytical results against numerical quantum trajectories, identifying a dynamical transition in the distribution profiles. This single-qubit setting provides a controlled and exact benchmark for analytical methods that can be extended to more complex monitored systems.

\section*{Model and Methods}

\begin{figure*}
\centering
\includegraphics[width=0.55\linewidth]{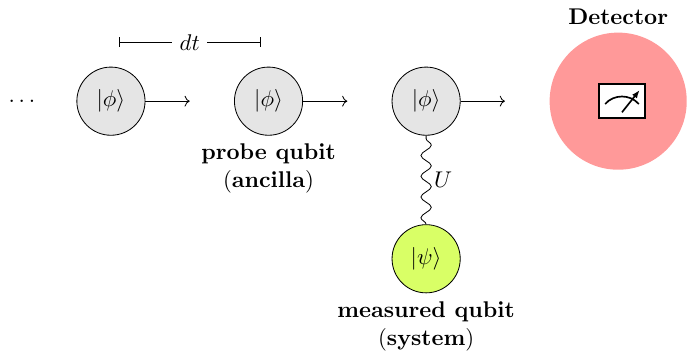}
\caption{Collisional implementation of continuous monitoring. A system qubit $\mathcal{S}$ interacts during a short time interval $dt$ with one ancillary probe prepared in the state $\ket{\phi}$ through the weak unitary $U=Z_{\mathcal{S}}(\theta)U_{\mathrm{CNOT}}(\theta)$. The ancilla is then projectively measured and discarded, so that each collision extracts only partial information about the system. In the scaling limit $\theta=\sqrt{\eta dt}\ll1$, the sequence of weak updates converges to a Markovian conditioned evolution, yielding the stochastic master equation used throughout the paper.}
     \label{fig: indirect measurement}
 \end{figure*}

To derive the continuous-time evolution of the system, we employ a collisional model framework \cite{Ciccarello_2022}, as illustrated in Fig.~\ref{fig: indirect measurement}. In this approach, a system qubit $\mathcal{S}$ interacts sequentially with a stream of ancillary probe qubits, each prepared in the state $\ket{\phi} = (\ket{0} - i\ket{1})/\sqrt{2}$. The interaction is governed by the unitary operator $U = Z_\mathcal{S}(\theta) U_{\text{CNOT}}(\theta)$, where $U_{\text{CNOT}}(\theta) = \cos\theta - i\,\text{CNOT}\sin\theta$ and $Z_\mathcal{S}(\theta) = e^{i\theta\sigma_z/2} \otimes \hat{1}$ is a local phase correction that ensures the diffusion is aligned with the measurement basis. The parameter $\theta$ modulates the coupling strength. 

We define the continuous monitoring limit by setting $\theta = \sqrt{\eta dt} \ll 1$, where $\eta$ is the measurement rate and $dt$ is the infinitesimal time interval between collisions. Following the interaction, a projective measurement is performed on the ancilla in the computational basis $\{\ket{0}, \ket{1}\}$. By expanding the resulting system state updates to $\mathcal{O}(dt)$ and mapping the discrete measurement outcomes to a Wiener process $dW$ (satisfying $dW^2 = \eta dt$), we obtain the Stochastic Master Equation (SME) describing the conditioned dynamics of the density matrix $\rho$:

\begin{equation}\label{eq:SME 1 qubit}
\begin{split}
d\rho = &-\frac{\eta}{2}[L,[L,\rho_t]] dt + (\rho_t L+L\rho_t-2\ev*{L}\rho_t)dW,
\end{split}
\end{equation}

where the Lindblad operator is $L = \ket{1}\bra{1}$ and $\ev{L} = \tr(L\rho_t)$. The deterministic double-commutator term represents the ensemble-averaged decoherence resulting from the interaction, while the stochastic term represents the gain of information provided by the measurement record. This evolution drives individual quantum trajectories toward pure states, while the unconditioned (averaged) state approaches a maximally mixed configuration. For systems with internal dynamics, the unitary contribution $-i/\hbar [H, \rho]dt$ is included linearly in Eq.~\eqref{eq:SME 1 qubit}.

When monitoring a qubit initialized in the maximally mixed state $\rho_0 = \hat{1}/2$, the absence of internal Hamiltonian dynamics ensures that no off-diagonal terms are generated. The state $\rho_t$ remains diagonal in the measurement basis and can be parameterized by a single stochastic variable $q(t) \in [-1, 1]$ as:
\begin{equation}\label{eq: rho as function of q}
\rho(q) = \frac{1 + q\sigma_z}{2},
\end{equation}
where $\sigma_z=\mathrm{diag}(1,-1)$ is the Pauli matrix in the $z$ direction and $q=1$ ($q=-1$) corresponds to the pure state $\ket{0}$ ($\ket{1}$). Substituting Eq.~\eqref{eq: rho as function of q} into the SME \eqref{eq:SME 1 qubit}, the evolution reduces to a multiplicative Langevin equation in the Itô sense:
\begin{equation}\label{eq: langevin equation for q}
dq = (1-q^2)dW,
\end{equation}
where $\ev{dW} = 0$ and $dW^2 = \eta dt$. The term $(1-q^2)$ acts as a state-dependent noise amplitude that vanishes at the boundaries $q = \pm 1$, reflecting the purification of the state toward the measurement eigenstates.

\begin{figure}[h]
    \centering
    \includegraphics[width=0.45\textwidth]{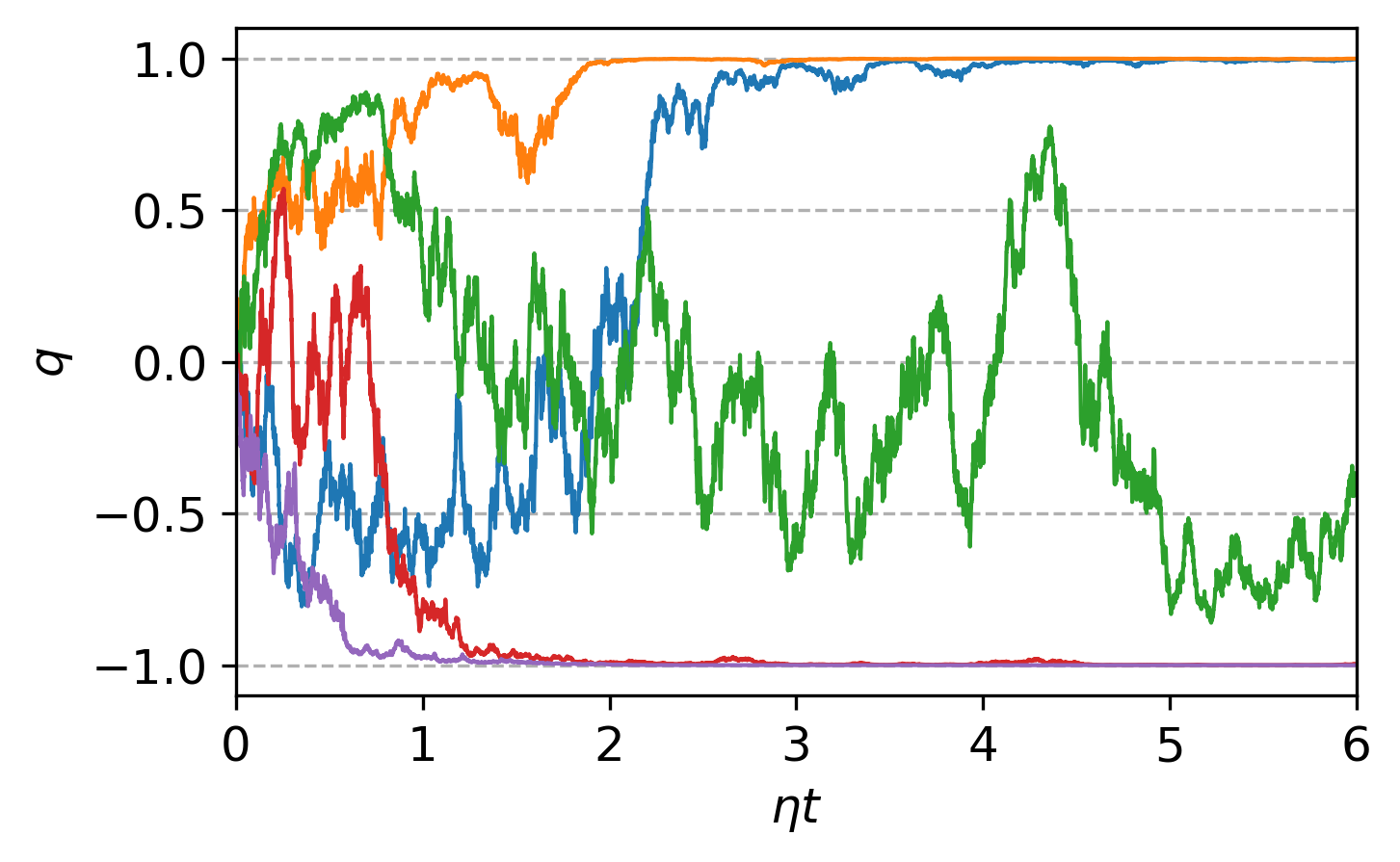}
    \caption{Individual trajectories $q(t)$ versus $\eta t$ generated from Eq.\eqref{num}. Notice that, in the present time scale, $q$ is naturally bounded in $\pm 1$.}
    \label{fig2}
\end{figure}

For the numerical calculation of quantum trajectories and the subsequent generation of histograms, we employ the Euler--Maruyama method \cite{Maruyama1955}. The time interval is discretized into steps of size $\Delta t$, with the update rule:
\begin{equation}
q_{n+1} = q_n + (1 - q_n^2) \Delta W_n,
\label{num}
\end{equation}
where $q_0=0$ (mixed state) and $\Delta W_n$ are independent Gaussian random variables with average zero and variance $\eta \Delta t$. For the simulations we set $\Delta t = 10^{-3}$ and  we average over $10^5$ independent trajectories to characterize the statistical properties of the purification process. Let us emphasize that no explicit boundary constraint was employed during the simulation. With $\Delta t = 10^{-3}$, the drift term $(1-q^2)$ becomes extremely small near the boundaries, and none of the $10^5$ trajectories exceeded $|q|=1$ within the simulated times (see Fig.\ref{fig2}).

\section*{Results}

Using It\^{o} calculus, the change of variables $Q=\mathrm{atanh}\,q$ obeys $dQ=Q' dq+\frac{1}{2}Q'' (dq)^2$, with $dW^2=\eta dt$. This gives
\begin{equation}
    dQ=\eta\, \tanh Q\,dt+dW\,.
    \label{itoQ}
\end{equation}
This process is known as the hyperbolic Ornstein--Uhlenbeck, or Bene\v{s}, process \cite{Benes1981,SarkkaSolin2019}. The corresponding It\^{o} Fokker--Planck equation is

\begin{equation}
    \partial_t P_Q=-\partial_Q [\eta \tanh Q P_Q]+\frac{\eta}{2}\partial_Q^2 P_Q\,.
    \label{FPKPQ}
\end{equation}

This process has no equilibrium distribution, since $P_{Q,eq}=\cosh^2 Q$ is non-normalizable for $Q\in (-\infty, \infty)$

\subsection*{The Onsager--Machlup approach and exact solution}

We now consider the path probability distribution $P(Q,t|Q_0,t_0)$ for a stochastic trajectory $Q$ governed by 
\begin{equation}
    P(Q,t|Q_0,t_0)\equiv P(Q,t)=
    \int\limits_{Q(0)=0}^{Q(t)=Q} \mathcal{D}Q\, e^{-S[Q]},
\end{equation}
where $Q_0=0$, $t_0=0$, and
\begin{equation*}
    S[Q]=\int_{0}^{t}dt'\left[\frac{(\dot Q(t')-\eta\tanh Q(t'))^2}{2\eta}+\eta\frac{\mathrm{sech^2} Q(t')}{2}\right],
\end{equation*}
is the action.  The equation above can be further simplified by the identity $\mathrm{sech}^2 Q=1-\tanh^2 Q$ and expanding the square term $(\dot Q-\eta\tanh Q)^2$, resulting in:
\begin{equation}\label{eq: S[Q]}
    S[Q]=\int_0^t dt'\left(\frac{\dot Q^2}{2\eta}-\dot Q\tanh Q\right).
\end{equation}
The term independent of $Q$ has been absorbed into the prefactor. Eq.~\eqref{eq: S[Q]} therefore defines an action that encodes the nontrivial dynamics in the effective position- and velocity-dependent potential
\begin{equation*}
    V(Q,\dot Q)=-\dot Q\tanh Q .
\end{equation*}

The key observation is that $\dot Q\tanh Q$ is a total time derivative:
\begin{equation*}
    \int_0^t \dot Q(t')\tanh Q(t') \,dt' = \int_{0}^{Q} \tanh Q' \,dQ' = \ln\cosh Q\,,
\end{equation*}
where $Q(0)=0$. This structure is closely related to Girsanov's theorem: the drifted diffusion in Eq.~\eqref{itoQ} has the same diffusion coefficient as a Wiener process, and its path measure is absolutely continuous with respect to the Wiener measure, with a Radon--Nikodym weight determined by the drift \cite{Girsanov1960}. In the present case, that weight is precisely the boundary factor $\exp[\ln\cosh Q]$ together with the normalization term $\exp[-\eta t/2]$. Consequently, the action reduces to a Gaussian Wiener action plus a boundary contribution, yielding the probability
\begin{equation}
    P_Q(Q,t)=\frac{1}{\sqrt{2\pi\eta t}} \exp\!\left(-\frac{Q^2}{2\eta t}+\ln\cosh Q-\frac{\eta t}{2}\right).
    \label{PQ}
\end{equation}
No saddle-point or semiclassical approximation has been made; the Onsager--Machlup path integral is evaluated exactly because the Girsanov weight is integrable as a boundary term.

We stress that the midpoint discretization used to construct the Onsager--Machlup measure does not redefine the stochastic process. The process is fixed by the It\^{o} SDE in Eq.~\eqref{itoQ}, or equivalently by the Fokker--Planck equation in Eq.~\eqref{FPKPQ}. Because the nonlinear transformation $q\mapsto Q=\operatorname{atanh}q$ was performed using It\^{o}'s lemma, the resulting SDE for $Q$ has additive noise and a constant diffusion coefficient. For such additive-noise processes, the It\^{o} and Stratonovich SDEs have identical drift and therefore generate the same transition probability. In the path-integral representation, however, the choice of time slicing affects the short-time measure. With the midpoint (Stratonovich) discretization, this effect manifests as the Jacobian term $\eta\frac{\mathrm{sech^2} Q}{2}$ in the Onsager--Machlup action. Including this term ensures that the midpoint path integral reproduces exactly the propagator of the It\^{o} Fokker--Planck equation, rather than a different stochastic dynamics.

Because the action is quadratic in $\dot Q$ and the nonlinear contribution depends only on the endpoint through $\ln\cosh Q$, the path integral separates into an endpoint-dependent part and a Gaussian integral over fluctuations. To make this separation explicit, write each trajectory with fixed final value $Q(t)=\Omega t$ as $Q(t')=\Omega t'+\xi(t')$, where $\xi(0)=\xi(t)=0$. The variable $\Omega=Q(t)/t=\frac{1}{t}\int_0^t\dot Q(t')\,dt'$ is therefore the time-averaged velocity, while $\xi$ contains only fluctuations that leave the endpoints fixed. Because these fluctuations enter only through the quadratic term $\int_0^t\dot\xi^2\,dt'$, the Gaussian integral over $\xi$ is independent of $\Omega$ and factors out. It therefore cancels when computing normalized averages. Hence the exact probability distribution for $\Omega$ is
\begin{equation}
    P_\Omega(\Omega,t)=\frac{e^{-S[Q_\Omega]}}{\int_{-\infty}^{\infty} d\Omega\, e^{-S[Q_\Omega]}},
\end{equation}
with $S[Q_\Omega] = \frac{\Omega^2 t}{2\eta} - \ln\cosh(\Omega t)$ (the additive constant $\eta t/2$ from the action is absorbed into the normalization). Explicitly,
\begin{align}\label{eq: P(omega,t)}
    P_\Omega(\Omega,t)&=\frac{e^{-\frac{\Omega^2t}{2\eta}+\ln\cosh\Omega t}}{\int d\Omega\,e^{-\frac{\Omega^2t}{2\eta}+\ln\cosh\Omega t}}\nonumber\\
    &=\left(\frac{ t}{2\pi \eta }\right)^{1/2}\exp\left(-\frac{\Omega^2 t}{2\eta}+\ln\cosh{\Omega t}-\frac{\eta t}{2} \right).
\end{align}

Thus, the purification dynamics can be interpreted exactly as follows: the trajectory first selects an effective purification rate $\Omega$ from the distribution $P_\Omega(\Omega,t)$; conditioned on this value, the state parameter follows the deterministic curve $q(t)=\tanh(\Omega t)$. Positive and negative values of $\Omega$ correspond to purification toward the two orthogonal measurement eigenstates.

\begin{figure*}[htb]
    \centering
    \includegraphics[width=0.45\linewidth]{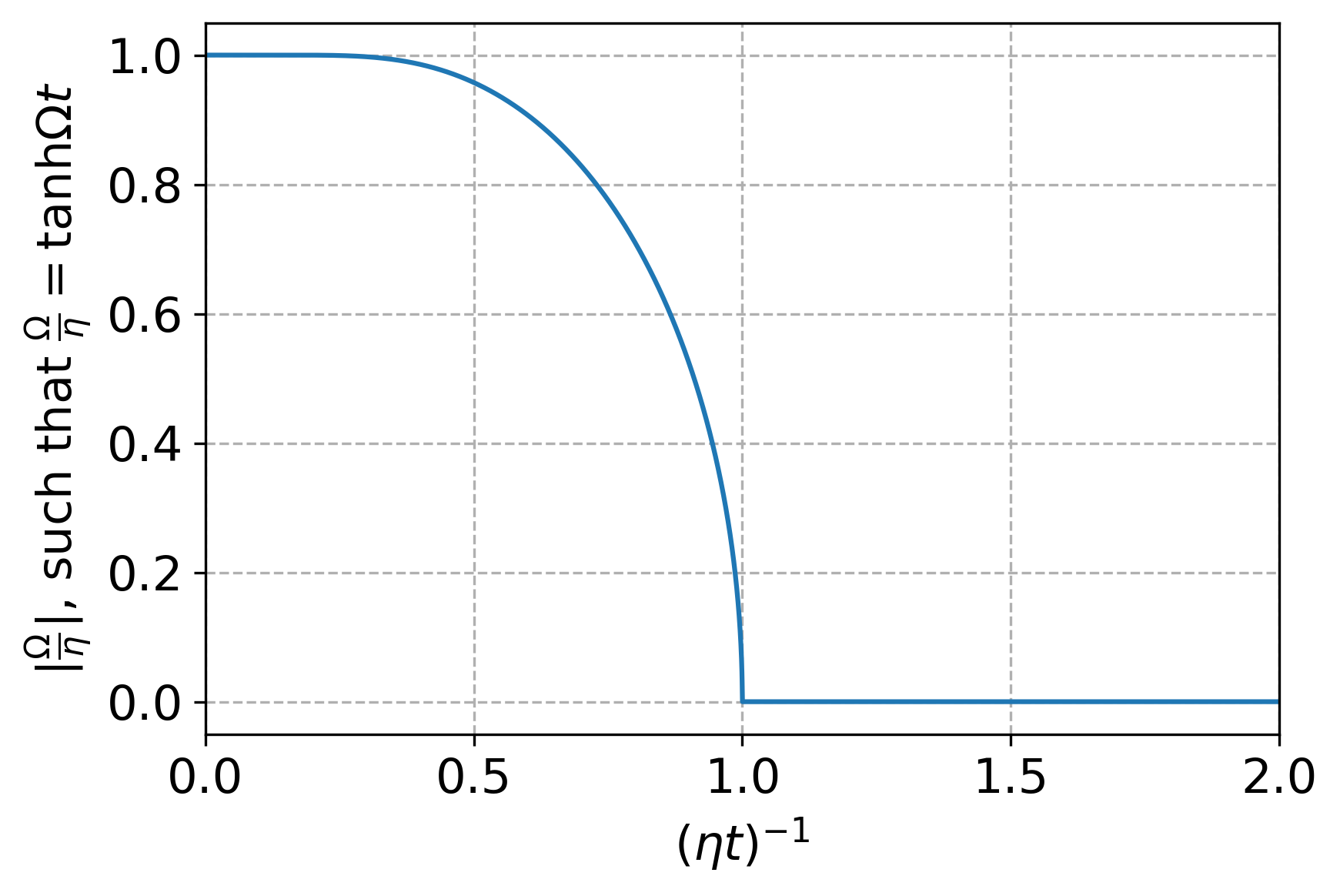}
    \includegraphics[width=0.45\linewidth]{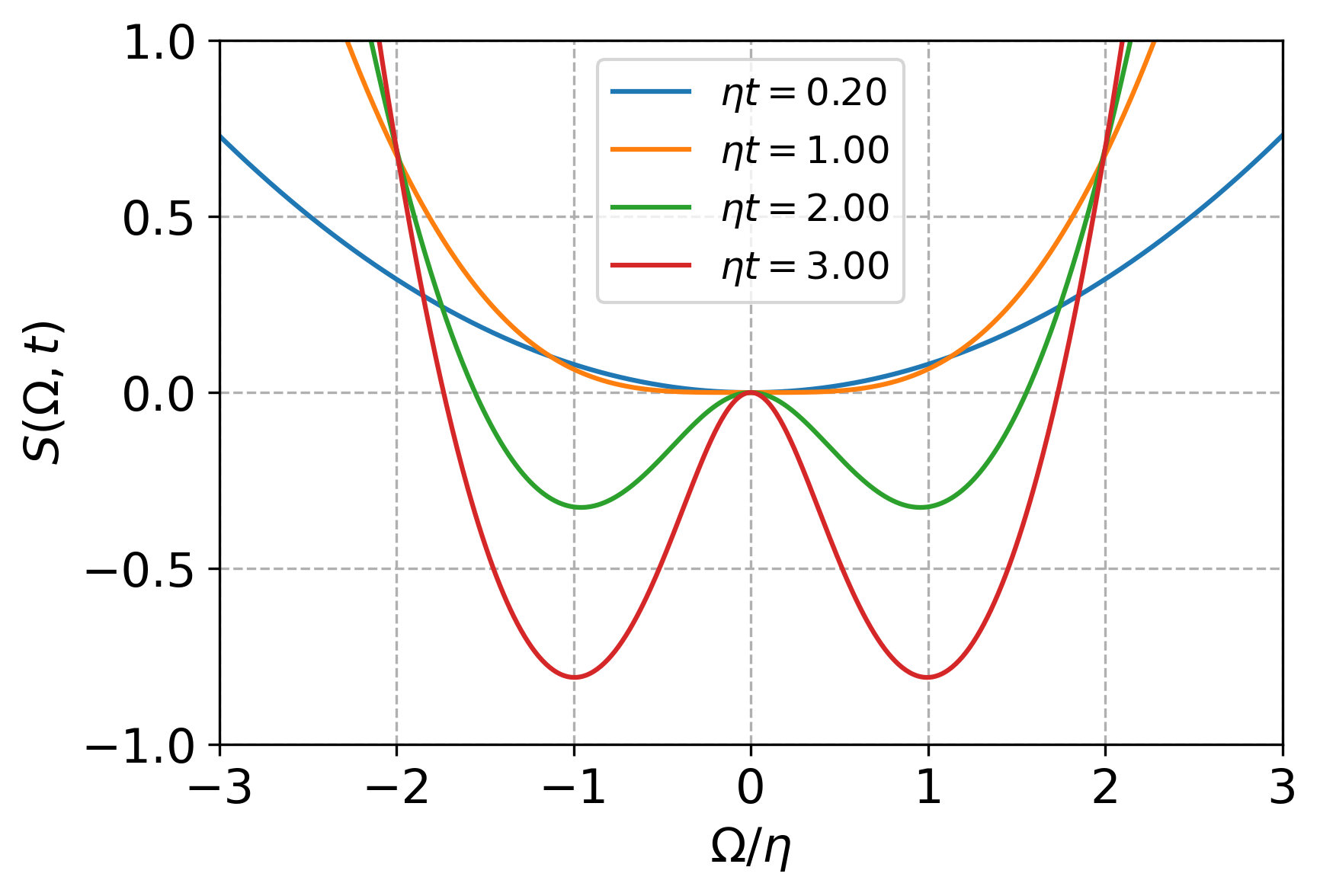}
    \caption{Left: solutions of the extremal condition $\Omega/\eta=\tanh(\Omega t)$, Eq.~\eqref{eq: non-linear eq}, plotted as a function of $(\eta t)^{-1}$. The nonzero branches appear for $\eta t>1$ and approach $\Omega=\pm\eta$ at long times. Right: action $S(\Omega,t)$ from Eq.~\eqref{eq: S(omega,t)} for increasing times. The central minimum at $\Omega=0$ becomes a local maximum and is replaced by two symmetric minima, providing the origin of the bimodal purification distribution.}
    \label{fig3}
\end{figure*}

The distribution $P_\Omega(\Omega,t)$ satisfies
\begin{equation*}
    P_\Omega(\Omega,t)\to\frac{\delta(\Omega-\eta)+\delta(\Omega+\eta)}{2},
\end{equation*}
in the asymptotic limit $\eta t\to\infty$. Therefore, at long times the distribution tends to a discrete bimodal form concentrated at $\Omega=\pm\eta$. Each peak corresponds to purification toward one of the two pure states, consistent with $q=\tanh(\Omega t)$. The two peaks emerge at a threshold time $t=t^*$, coinciding with the appearance of nontrivial minima in the classical action
\begin{equation}\label{eq: S(omega,t)}
    S(\Omega,t)=\frac{\Omega^2 t}{2\eta}-\ln\cosh\Omega t,
\end{equation}
for $\Omega\neq 0$. More precisely, the extrema satisfy
\begin{equation*}
    \left(\frac{\partial S}{\partial \Omega}\right)_t=\frac{\Omega t}{\eta}-t\tanh(\Omega t)=0.
\end{equation*}
Equivalently,
\begin{equation}\label{eq: non-linear eq}
    \frac{\Omega}{\eta}=\tanh(\Omega t).
\end{equation}
In the limit $t\to\infty$, Eq.~\eqref{eq: non-linear eq} admits the solutions $\Omega=\pm\eta$. For finite time, the corresponding extrema are shown in Fig.~\ref{fig3}. This figure translates the emergence of bimodality in $P_\Omega$ into the geometry of the action. The left panel tracks the solutions of the extremal equation as the inverse time $(\eta t)^{-1}$ is varied: above the threshold, only the symmetric solution at $\Omega=0$ is selected, whereas below the threshold two nonzero branches appear. The right panel shows the same mechanism directly in the action landscape, where the single central minimum progressively gives way to two symmetric minima. Thus, Fig.~\ref{fig3} identifies the onset of purification with a change in the dominant paths contributing to the Onsager--Machlup integral.

The threshold therefore occurs at $t^*=\eta^{-1}$. Figure~\ref{fig3} shows the crossover in the  structure of the action: the dominant contribution shifts from the central minimum at $\Omega=0$ to two symmetric nonzero minima. Correspondingly, the average $\ev{|\Omega|}_\Omega$ grows continuously from $0$ toward $\eta$ as time increases, with $\ev{...}_{\Omega}\propto\int d\Omega\, e^{-S(\Omega,t)}(...).$ In this sense, $\Omega$ behaves as an effective measurement rate. For $\eta t\leq 1$, the dynamics are dominated by diffusive fluctuations around the mixed state. For $\eta t>1$, the nonzero minima dominate the path weight, so individual trajectories are increasingly biased toward one of the two measurement outcomes. 

Having characterized the effective-rate distribution, we now transform back to physical observables. Since $q=\tanh(\Omega t)$, the probability density of the state coordinate follows directly from the change of variables $\Omega=\operatorname{atanh}(q)/t$:
\begin{align}\label{eq: P(q,t)}
    P_q(q,t)&=\frac{P_\Omega\!\bigl(\Omega=\mathrm{atanh}(q)/t,t\bigr)}
    {t(1-q^2)}\nonumber\\
    &=\frac{1}{\sqrt{2\pi\eta t}\,(1-q^2)}
    \exp\biggl[-\frac{\mathrm{atanh}^2(q)}{2\eta t}
    \nonumber\\
    &\hspace{1.9cm}
    +\ln\cosh\!\bigl(\mathrm{atanh}(q)\bigr)-\frac{\eta t}{2}\biggr].
\end{align}

The purity is an even function of $q$,
\begin{equation}
    \tau(q)=\frac{1+q^2}{2}\,,
    \label{purity}
\end{equation}
with $1/2\leq\tau\leq1$. Since the two values $q=\pm\sqrt{2\tau-1}$ correspond to the same purity, the associated probability density is
\begin{align}\label{eq: P(tau,t)}
    P_\tau(\tau,t)
    &=2\frac{P_q\!\bigl(q=\sqrt{2\tau-1},t\bigr)}
    {\sqrt{2\tau-1}},\nonumber\\
    &=\frac{\exp[-\eta t/2]}
    {\sqrt{2\pi\eta t}\,(1-\tau)\sqrt{2\tau-1}}
    \nonumber\\
    &\quad\times
    \exp\biggl[-\frac{\mathrm{atanh}^2\!\bigl(\sqrt{2\tau-1}\bigr)}
    {2\eta t}
    -\frac{1}{2}\ln\!\bigl[2(1-\tau)\bigr]\biggr].
\end{align}
This completes the chain of transformations $P_\Omega\to P_q\to P_\tau$: the effective-rate distribution determines the full statistics of the state coordinate and, consequently, of the purity.

The mean purity provides a compact scalar measure of this distribution. Using $q_\Omega=\tanh(\Omega t)$ in Eq.~\eqref{purity},
\begin{equation*}
    \ev\tau=\int_{-\infty}^{\infty}d\Omega\, P_\Omega(\Omega,t)\tau(q_\Omega),
\end{equation*}
one obtains
\begin{align}\label{eq: tau mean stratonovich}
    \ev{\tau}=\frac{1}{2}+\frac{1}{2}\left(\frac{ t}{2\pi \eta}\right)^{1/2}
    \int_{-\infty}^{\infty} d\Omega\,&
    e^{-\frac{\Omega^2 t}{2\eta}+\ln\cosh{\Omega t}-\frac{\eta t}{2}}\nonumber\\
    &\times\tanh^2(\Omega t).
\end{align}

\begin{figure*}[htb]
    \centering
    \includegraphics[width=0.45\linewidth]{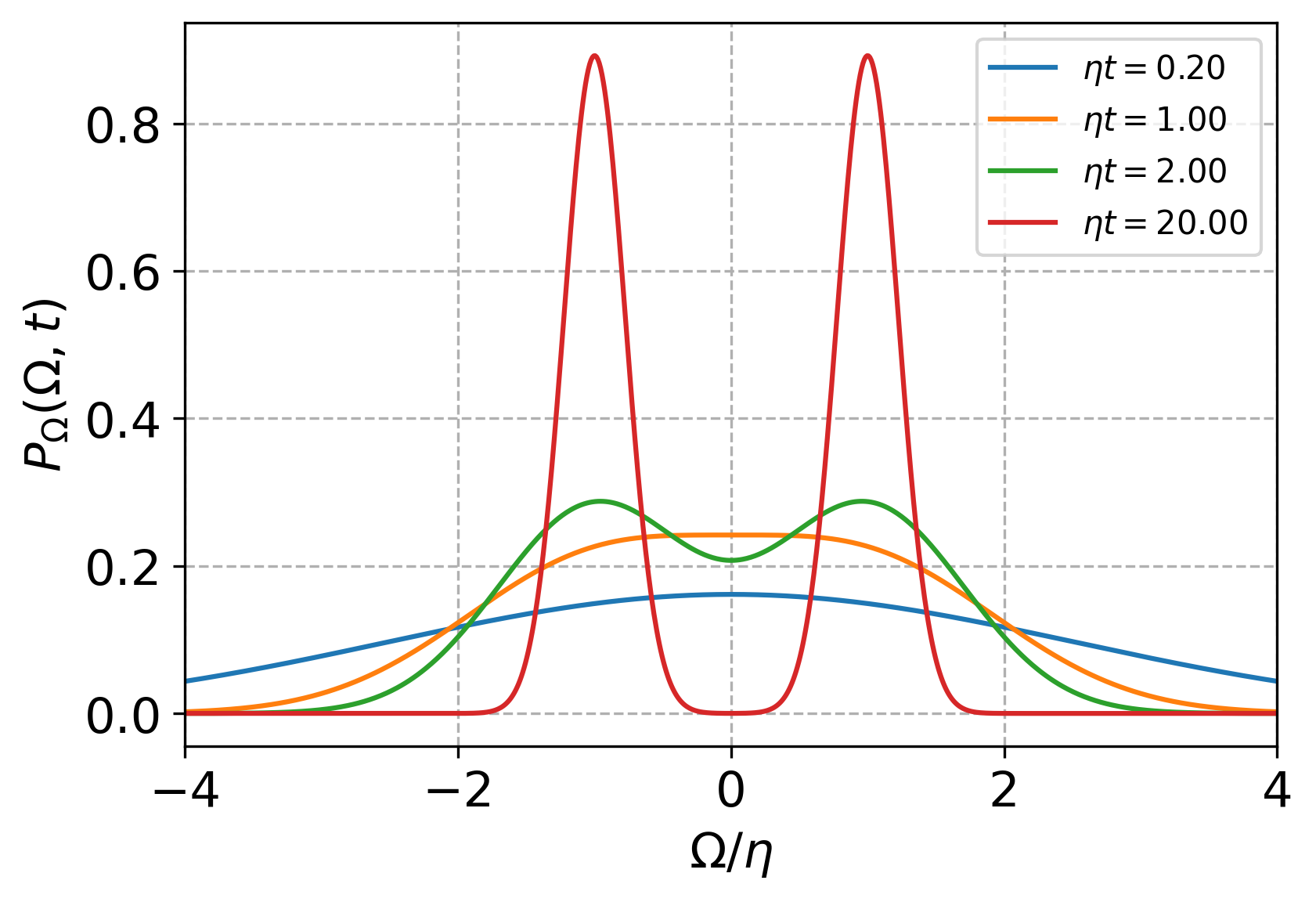}
    \includegraphics[width=0.45\linewidth]{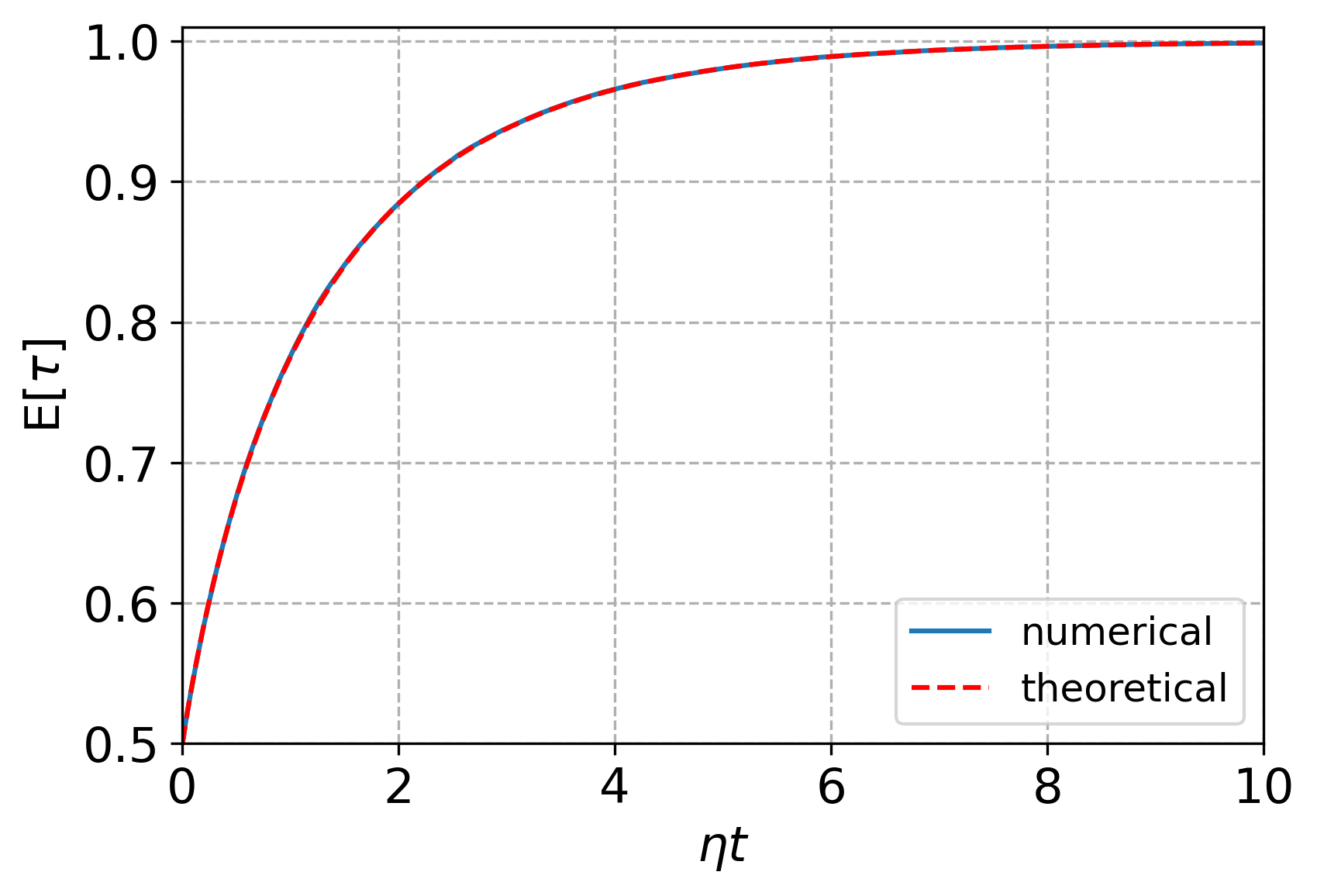}    
    \caption{Left: probability density $P_\Omega(\Omega,t)$ versus $\Omega/\eta$ from Eq.~\eqref{eq: P(omega,t)} for increasing values of the dimensionless time $\eta t$. The initially single-peaked distribution broadens and evolves toward two dominant branches at finite $\Omega$, signaling the onset of measurement-selected trajectories. Right: mean purity $\mathrm{E}[\tau]=\ev{\tau}$ obtained by averaging $\tau(q)$ over $100,000$ Euler--Maruyama trajectories of Eq.~\eqref{eq: langevin equation for q} (solid blue line) and from the analytical expression in Eq.~\eqref{eq: tau mean stratonovich} (red dashed line) versus $\eta t$. }
    \label{fig4}
\end{figure*}

Figure~\ref{fig4} summarizes the effective-rate distribution and compares the analytical mean purity with numerical trajectories.

\begin{figure*}[htb]
    \centering
    \includegraphics[width=\linewidth]{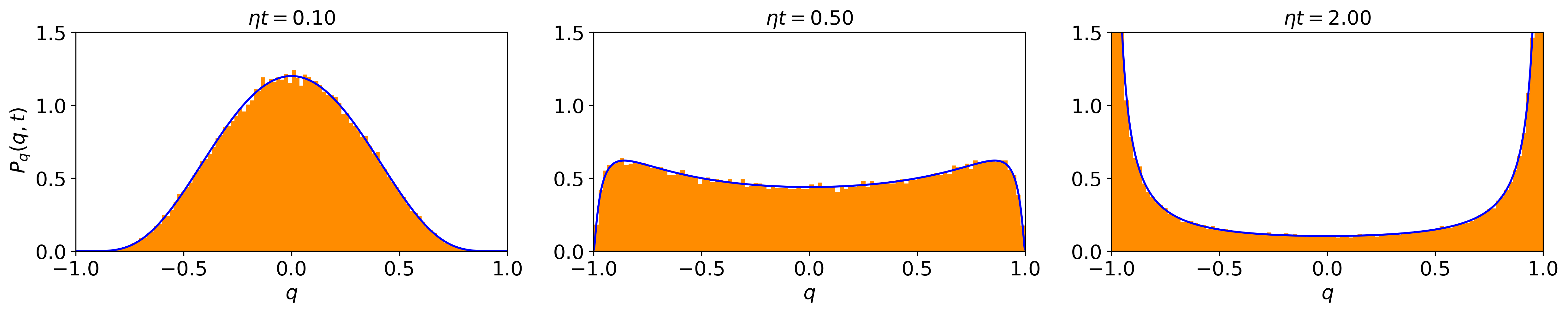}
    \includegraphics[width=\linewidth]{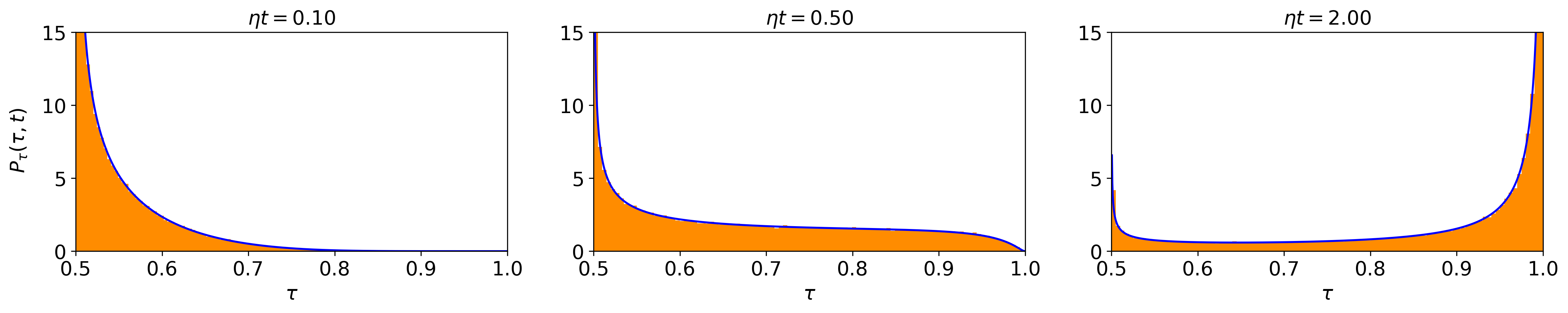}
    \caption{Full trajectory distributions for the monitored qubit. Top: probability density $P_q(q,t)$ for $\eta t=0.1,0.5,2.0$, comparing numerical trajectory histograms (orange) with the analytical result of Eq.~\eqref{eq: P(q,t)} (solid blue line). Bottom: corresponding purity distribution $P_\tau(\tau,t)$ obtained from Eq.~\eqref{eq: P(tau,t)}. At short times the state remains concentrated near the maximally mixed value $q=0$ and $\tau=1/2$, whereas at later times the weight shifts toward $|q|\simeq1$ and $\tau\simeq1$, demonstrating measurement-induced purification.}
    \label{fig5}
\end{figure*}

Figure~\ref{fig5} compares the analytical densities $P_q(q,t)$ and $P_\tau(\tau,t)$ with numerical trajectory histograms. This final comparison closes the connection between the auxiliary rate variable and directly observable quantities. In the top panels, $P_q(q,t)$ shows how the state coordinate moves away from the maximally mixed point $q=0$ and accumulates probability near the measurement eigenstates $q=\pm1$. In the bottom panels, the same evolution is compressed into the purity variable: probability weight is transferred from $\tau=1/2$ toward $\tau=1$. The agreement between histograms and analytical curves demonstrates that the full trajectory statistics, not only the mean purity, are captured by the path-integral solution.

\subsection*{Compatibility with the Fokker--Planck equation}

The Fokker--Planck formulation provides an independent consistency check for the probability distribution obtained from the Onsager--Machlup construction. Nonlinear drift terms usually prevent simple closed-form propagators, and their analysis often requires spectral decompositions, similarity transformations, or numerical methods \cite{Risken1989,gardiner1985handbook,van_kampen}. A closely related weak-measurement problem was discussed in Ref.~\cite{Patel2015}, but without deriving the explicit transition probability. The monitored-qubit dynamics considered here is special because the drift in Eq.~\eqref{itoQ} is proportional to the logarithmic derivative of $\cosh Q$, which makes the propagator analytically tractable.

To verify compatibility with the Fokker--Planck equation in Eq. \eqref{FPKPQ}, we compare the time derivative of Eq.~\eqref{PQ} with the sum of its drift and diffusion contributions. The time derivative is
\begin{equation}\label{eq: dt PQ}
    \partial_t P_Q
    =\left(\frac{Q^2}{2\eta t^2}-\frac{1}{2t}-\frac{\eta}{2}\right)P_Q.
\end{equation}
The drift term gives
\begin{equation}\label{eq: drift PQ}
    -\eta\partial_Q\!\left[\tanh Q\,P_Q\right]
    =\left(\frac{Q\tanh Q}{t}-\eta\right)P_Q,
\end{equation}
while the diffusion term gives
\begin{equation}\label{eq: diffusion PQ}
    \frac{\eta}{2}\partial_Q^2P_Q
    =\left(-\frac{Q\tanh Q}{t}+\frac{\eta}{2}
    -\frac{1}{2t}+\frac{Q^2}{2\eta t^2}\right)P_Q.
\end{equation}
Adding Eqs.~\eqref{eq: drift PQ} and \eqref{eq: diffusion PQ} yields Eq.~\eqref{eq: dt PQ}, proving that Eq.~\eqref{PQ} solves the Fokker--Planck equation exactly.

This exact compatibility confirms that the Onsager--Machlup path-integral result is not an approximation but reproduces the full transition probability. Moreover, it highlights why this model is a valuable benchmark: once Hamiltonian dynamics, additional measurement channels, or many-body scrambling terms are introduced, the Fokker--Planck equation will generally lose this solvable structure. In that more general setting, the path-integral formulation remains a systematic starting point for controlled analytical approximations.

\section*{Final remarks}

We have analyzed the purification of a single qubit under continuous monitoring using a collisional model in which the system interacts sequentially with identically prepared ancillary qubits. After tracing out the measured ancillas, the conditioned dynamics reduces to a stochastic master equation. For initially mixed states diagonal in the measurement basis, this dynamics is fully described by a single variable $q(t)$, whose evolution is governed by the multiplicative-noise Langevin equation in Eq.~\eqref{eq: langevin equation for q}. This reduction makes the monitored qubit a minimal but nontrivial setting in which the full statistics of purification can be studied analytically.

The main result of this work is the exact probability distribution for the monitored trajectories. By introducing the nonlinear variable $Q=\operatorname{atanh}(q)$, the stochastic dynamics takes a form that can be treated with the Onsager--Machlup path-integral formalism. The resulting distribution $P_\Omega(\Omega,t)$ leads directly to closed expressions for $P_q(q,t)$ and for the purity distribution $P_\tau(\tau,t)$. These analytical predictions agree with numerical trajectories generated by the Euler--Maruyama scheme, confirming that the path-integral construction captures the relevant stochastic dynamics of the continuously monitored qubit.

A central physical consequence is the emergence of a bimodal structure in the purification-related distributions. At short times, the dynamics is dominated by diffusive fluctuations and the state remains broadly distributed around the maximally mixed configuration. At longer times, the distribution develops two peaks associated with purification toward the two measurement eigenstates \cite{Patel2015}. This crossover is encoded in the effective rate variable $\Omega$ and in the structure of the action $S(\Omega,t)$, which develops nonzero extrema when measurement-induced purification becomes dominant.

The compatibility with the Fokker--Planck equation further strengthens the result. Although nonlinear Fokker--Planck equations rarely admit closed-form propagators, the present problem has a special solvable structure. The exact solution obtained in Eq.~\eqref{PQ} verifies the path-integral prediction and clarifies why this single-qubit model is an important benchmark: it provides a case where trajectory simulations, Fokker--Planck analysis, and Onsager--Machlup methods can be compared without uncontrolled approximations.

These results establish a useful analytical reference point for monitored quantum dynamics. While a single qubit cannot display a genuine many-body measurement-induced phase transition, it isolates the statistical mechanism by which continuous measurement extracts information and drives purification. The framework developed here can therefore serve as a starting point for more complex settings, including monitored systems with Hamiltonian dynamics, additional measurement channels, feedback, or coupling to scrambling degrees of freedom.

\section*{Acknowledgments}
We acknowledge fruitful discussions with T. Micklitz as well as partial financial support from  Conselho Nacional de Desenvolvimento Científico e Tecnológico (CNPq) , Fundação Carlos Chagas Filho de Amparo à Pesquisa do Estado do Rio de Janeiro (FAPERJ) and  Coordenação de Aperfeiçoamento de Pessoal de Nível Superior (CAPES) (Brazilian agencies).

\end{document}